\documentstyle[prd,aps,epsfig]{revtex}
\begin{document}
\draft
\title{Bounds on the electromagnetic interactions of excited
spin-3/2 leptons}
\author{R. Walsh\thanks{e-mail address: walsh@if.ufrj.br}~ and
A. J. Ramalho\thanks{e-mail address: ramalho@if.ufrj.br}}
\address{Instituto de F\'{\i}sica, Universidade Federal do Rio de Janeiro, 
Caixa Postal 68528, Ilha do Fund\~ao,\\
21945-970 Rio de Janeiro, RJ, Brazil}
% \begin{document}
\maketitle
\begin{abstract}
We discuss possible deviations from QED produced by a virtual excited spin-3/2 
lepton in the reaction $e^+e^-\longrightarrow 2\gamma$. Data recorded by the 
OPAL Collaboration at a c.m. energy $\sqrt{s} = 183\ GeV$ are used to 
establish bounds on the nonstandard-lepton mass and coupling strengths.
\end{abstract}

The success of the standard model in describing the existing
phenomenology of the electroweak and strong interactions is rather impressive.
Yet few theorists believe the standard model is a fully satisfactory theory of
fundamental interactions, since it leaves some important questions unanswered.
In view of the shortcomings of the standard model, a host of extended
models has been put forward, which predict the existence of new particles
and interactions. The search for the manifestations of this new physics is
a major task to be undertaken by the experimental groups at the present
and future colliders. Here we discuss possible effects of an excited
spin-3/2 lepton on two-photon production in $e^+e^-$ collisions. In the
literature exotic spin-3/2 particles have appeared in different contexts,
with their production rates and decay modes being analyzed in the
environments of $e^+e^-$, $ep$, $e\gamma$, $\gamma\gamma$ and $pp$ collisions 
\cite{lopes1,lopes2,kuhn}.
Supersymmetric theories are known to include supermultiplets with spin-3/2
particles. In supergravity gauge theories there are fundamental spin-3/2
fermions, the gravitinos, which can be endowed with typical quantum
numbers of the ordinary quarks and leptons. Spin-3/2 fermions are also
present in composite models \cite{lopes2,kuhn}, in which deviations from the 
standard model are due to an underlying substructure of quarks and leptons.

Field theories for interacting spin-3/2 particles are known to be
nonrenormalizable, violating unitarity at sufficiently high energies 
\cite{lopes2}. In order to parametrize the effects of a nonstandard spin-3/2 
lepton interacting with electrons and photons, we consider two effective 
interaction lagrangians
\begin{eqnarray}
{\cal L}^{(1)}_{int} & = & {e \over \Lambda} {\bar{\Psi}}^*_{\mu} 
{\gamma}_{\nu} 
(c_L\psi_L + c_R\psi_R) F^{\mu\nu} \mbox{ ,} \nonumber \\
{\cal L}^{(2)}_{int} & = & {e \over \Lambda^2} {\bar{\Psi}}^*_{\mu} 
{\sigma}_{\alpha\beta} (c_L\psi_L + c_R\psi_R)\partial^{\mu} 
F^{\alpha\beta} \mbox{ ,} \nonumber
\end{eqnarray} 
where $\Psi_{\mu}$ is a Rarita-Schwinger vector-spinor field representing
the excited spin-3/2 lepton, $\psi_{L,R}$ are definite-helicity Dirac
spinor fields corresponding to the electrons and $F^{\mu\nu}$ the
electromagnetic field strength. $\Lambda$ is a characteristic energy scale
around which effects of the new physics would become manifest. Both
lagrangians above are gauge invariant. It is important to point out that,
to avoid running into conflict with $(g-2)$ measurements of electrons and
muons, one must couple the spin-3/2 lepton exclusively to left-handed or
right-handed ordinary leptons \cite{kuhn}.

\begin{figure}
\begin{center}
\mbox{\epsfig{file=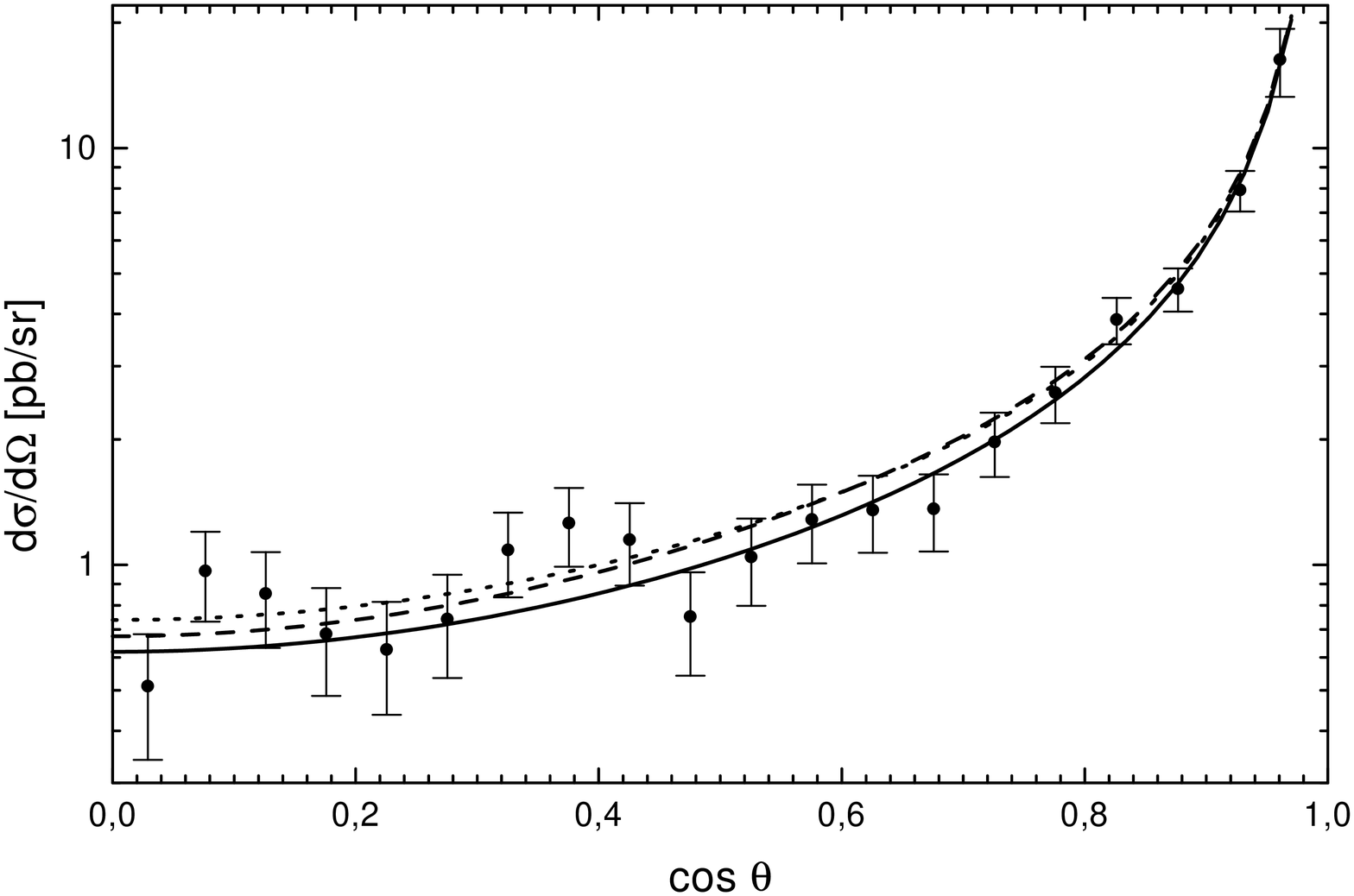,width=7.5cm,height=6cm}}
\end{center}
\caption{Angular distribution at $\sqrt s=183\ GeV$. The solid curve 
represents the QED prediction, whereas the dashed (dotted) curve shows the 
total angular spectrum in the presence of the nonstandard interaction 
${\cal L}_{int}^{(1)}$ (${\cal L}_{int}^{(2)}$) for an input mass 
$M_{3/2}=125\ GeV$ ($142\ GeV$). OPAL data are also shown for comparison.}
\label{fig1}
\end{figure}

The process $e^+e^- \longrightarrow 2\gamma$ is a very convenient
tool to search for physics beyond the standard model. The total and
differential cross sections can be measured  with precision at the LEP 
detectors \cite{opal,aleph}. We used data taken by the OPAL Collaboration 
\cite{opal} at a center-of-mass energy $\sqrt{s} = 183\ GeV$ and total 
integrated luminosity of $56.2\ pb^{-1}$ to obtain lower bounds on the mass 
scale $\Lambda$, as well as on the spin-3/2 excited-lepton mass $M_{3/2}$ and 
coupling strengths $c_{L,R}$. The calculation of the differential cross section 
for two-photon production was performed at tree level, taking into account the
nonstandard couplings specified by ${\cal L}^{(1)}_{int}$ and 
${\cal L}^{(2)}_{int}$. The resulting expressions are given by
$$
{d{\sigma}^{(i)} \over d\Omega} = \Big({d\sigma \over d\Omega}\Big)_{QED} +
{{\alpha}^2 \over 16s} \Big[F^{(i)}_{+}(c_L,c_R,x,y,s/\Lambda^2) +
F^{(i)}_{-}(c_L,c_R,x,y,s/\Lambda^2) \Big] \mbox{ , } i=1,2 \mbox{ ,}
$$
where $({d\sigma/ d\Omega})_{QED} = (\alpha^2/s){(1+x^2)/(1-x^2)}$ is the 
photon angular distribution expected from QED, $x \equiv
cos{\theta}$, $y \equiv 2M^2_{3/2}/s$, and the nonstandard corrections read
\pagebreak
\begin{table}
\caption{Coefficients $a_n(y)$ for the polynomials of the corrections 
$F_\pm^{(i)}$.}
\begin{center}
\begin{tabular}{cccccccccc}
 & $a_8$ & $a_7$ & $a_6$ & $a_5$ & $a_4$ & $a_3$ & $a_2$ & $a_1$ & $a_0$ \\ 
 \hline\hline
 $A_+^{(1)}$ & $0$ & $0$ & $-1$ & $2y+4$ & $-10y^2+2y$ & $-32y^2-28y$ & 
 $8y^3+48y^2$ & $-16y^3-32y^2$ & $80y^3+26y^2$ \\ 
 & & & & & $-5$ & & $+52y+5$ & $-38y-4$ & $+10y+1$ \\ \hline
 $A_-^{(1)}$ & $0$ & $0$ & $-1$ & $2y+4$ & $-10y^2-14y$ & $36y$ & 
 $-8y^3-48y^2$ & $16y^3+64y^2$ & $-80y^3-6y^2$ \\ 
 & & & & & $-5$ & & $-44y+5$ & $+26y-4$ & $-6y+1$ \\ \hline
$B_+^{(1)}$ & $0$ & $0$ & $0$ & $0$ & $3y+13$ & $0$ & $-4y^2-8y$ & $0$ & 
$40y^2+5y$ \\ 
 & & & & & & & $-14$ & & $+1$ \\ \hline
$B_-^{(1)}$ & $0$ & $0$ & $0$ & $0$ & $3y+5$ & $0$ & $4y^2+8y$ & $0$ & 
$-40y^2-11y$ \\ 
 & & & & & & & $+2$ & & $-7$ \\ \hline
$C^{(1)}$ & $0$ & $0$ & $0$ & $0$ & $-2$ & $y+8$ & $-11y-12$ & $7y+8$ & 
$3y-2$ \\ \hline 
$D^{(1)}$ & $0$ & $0$ & $0$ & $0$ & $-1$ & $-3y-2$ & $y$ & $3y+2$ & $-y+1$ 
\\ \hline\hline
 $A_+^{(2)}$ & $-1$ & $6$ & $-4y-14$ & $14$ & $12y^2+60y$ & $-120y^2-160y$ & 
 $72y^3+288y^2$ & $-144y^3-264y^2$ & $72y^3+84y^2$ \\ 
 & & & & & & $-14$ & $+180y+14$ & $-96y-6$ & $+20y+1$ \\ \hline
 $A_-^{(2)}$ & $1$ & $-6$ & $20y+14$ & $-96y-14$ & $84y^2+180y$ & 
 $-264y^2-160y$ & $72y^3+288y^2$ & $-144y^3-120y^2$ & $72y^3+12y^2$ \\ 
 & & & & & & $+14$ & $+60y-14$ & $+6$ & $-4y-1$ \\ \hline
$B^{(2)}$ & $0$ & $0$ & $-1$ & $0$ & $-6y+3$ & $0$ & $-9y^2-3$ & $0$ & 
$9y^2+6y$ \\ 
 & & & & & & & & & $+1$ \\ \hline
$C^{(2)}$ & $0$ & $0$ & $0$ & $-1$ & $3$ & $-2y-2$ & $6y-2$ & $-6y+3$ & 
$2y-1$ \\ \hline 
$D^{(2)}$ & $0$ & $0$ & $0$ & $0$ & $-2$ & $-4y-4$ & $-4y$ & $4y+4$ & 
$4y+2$
\end{tabular}
\end{center}
\label{table}
\end{table} 
\begin{eqnarray}
F_{\pm}^{(1)} & = & {s^2 \over \Lambda^4} 
{(c_R^2\pm c_L^2)^2\over 72y^2(1-y-x)} \left[ {A_\pm^{(1)}(x,y) 
\over (1-y-x)}+{2yB_\pm^{(1)}(x,y)\over (1+y+x)} \right] + {s \over 
\Lambda^2}{(c_R^2+c_L^2)\over 6y(1-x)} \left[ {C^{(1)}(x,y)\over 
(1-y-x)}+{D^{(1)}(x,y) \over (1+y+x)}\right] + (x \rightarrow -x)
\mbox{ ,} \nonumber \\
F_{\pm}^{(2)} & = & {s^4 \over \Lambda^8} 
{(c_R^2\pm c_L^2)^2\over 288y^2(1-y-x)} \left[ {A_\pm^{(2)}(x,y) 
\over (1-y-x)}+{4yB^{(2)}(x,y)\over (1+y+x)} \right] + {s^2 \over 
\Lambda^4}{c_R c_L \over 3y(1-x)} \left[ {C^{(2)}(x,y)\over 
(1-y-x)}+{D^{(2)}(x,y) \over (1+y+x)}\right] + (x \rightarrow -x) 
\mbox{ ,} \nonumber
\end{eqnarray}
where $A_\pm^{(i)}$, $B_\pm^{(1)}$, $B^{(2)}$, $C^{(i)}$ and $D^{(i)}$, 
$i=1,2$, are polynomials written in the form $\sum_n a_n(y)x^n$, with the
y-dependent coefficients $a_n(y)$ given in Table\ \ref{table}. Fig.\ 
\ref{fig1}  shows the
angular distributions $d\sigma^{(i)}/d\Omega$ at $\sqrt{s} = 183\ GeV$,
along with the corresponding prediction for QED and OPAL experimental data.
In line with OPAL experimental procedure, we consider the event
angle $\theta$ defined so that $cos{\theta}$ is positive, since the two
photons are identical, and an experimental cut $\cos \theta<0.97$. The 
compositeness scale $\Lambda$ was taken to be equal to the
exotic-lepton mass, with numerical values consistent with the $95\%$
confidence level lower bounds that we derived for each interaction, as 
discussed in the following.

\begin{figure}
\begin{minipage}{.470\linewidth}
\begin{center}
\mbox{\epsfig{file=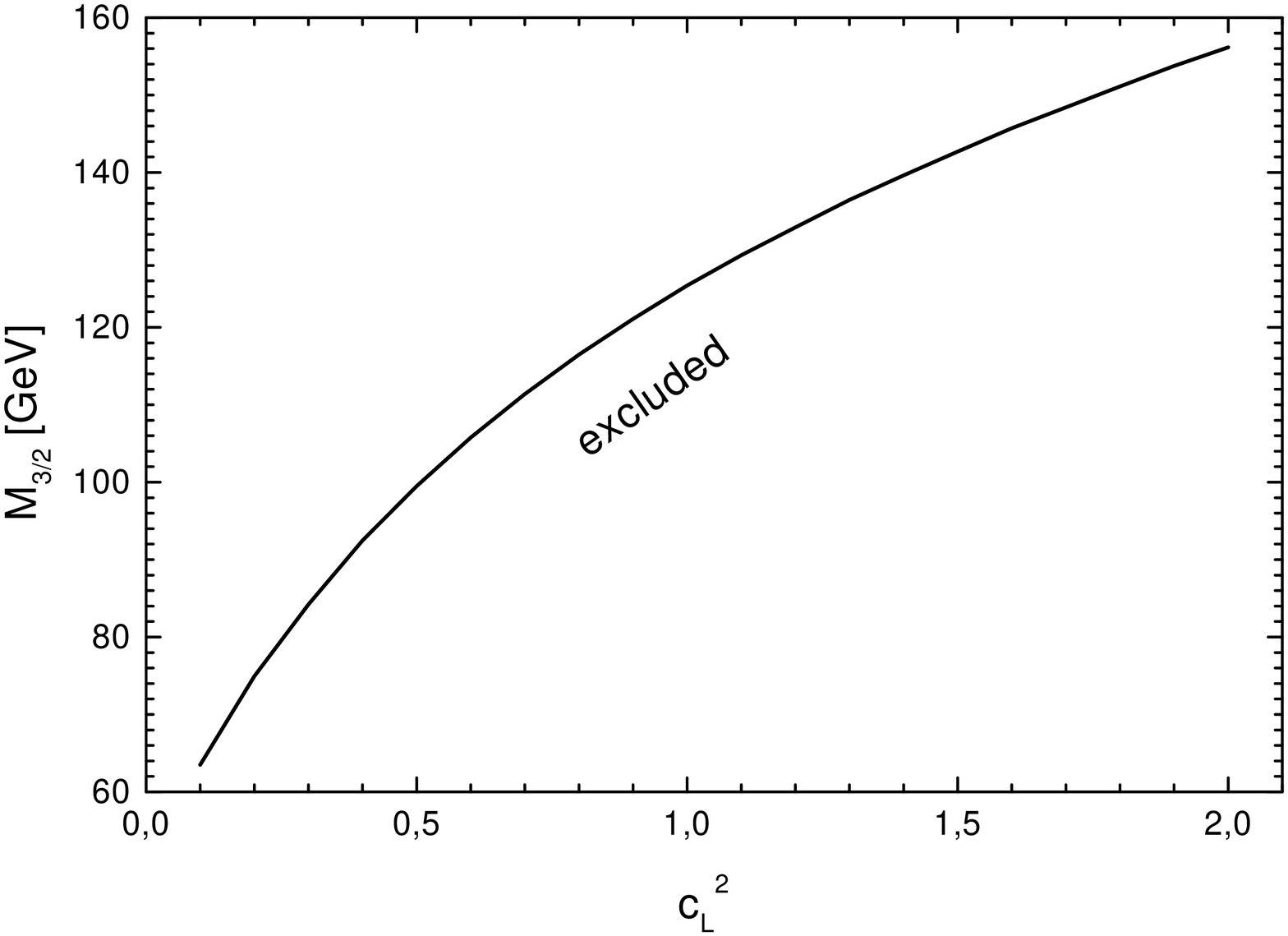,width=7.5cm,height=6cm}}
\end{center}
\caption{$95\%$ C.L. lower bound on the spin-3/2 lepton mass $M_{3/2}$ as a 
function of $c_L^2$ for interaction ${\cal L}_{int}^{(1)}$ and c.m. energy 
$\sqrt s=183\ GeV$.}
\label{fig2}
\end{minipage}
\nolinebreak
\hskip 0.5cm
\begin{minipage}{.470\linewidth}
\vskip -24pt
\begin{center}
\mbox{\epsfig{file=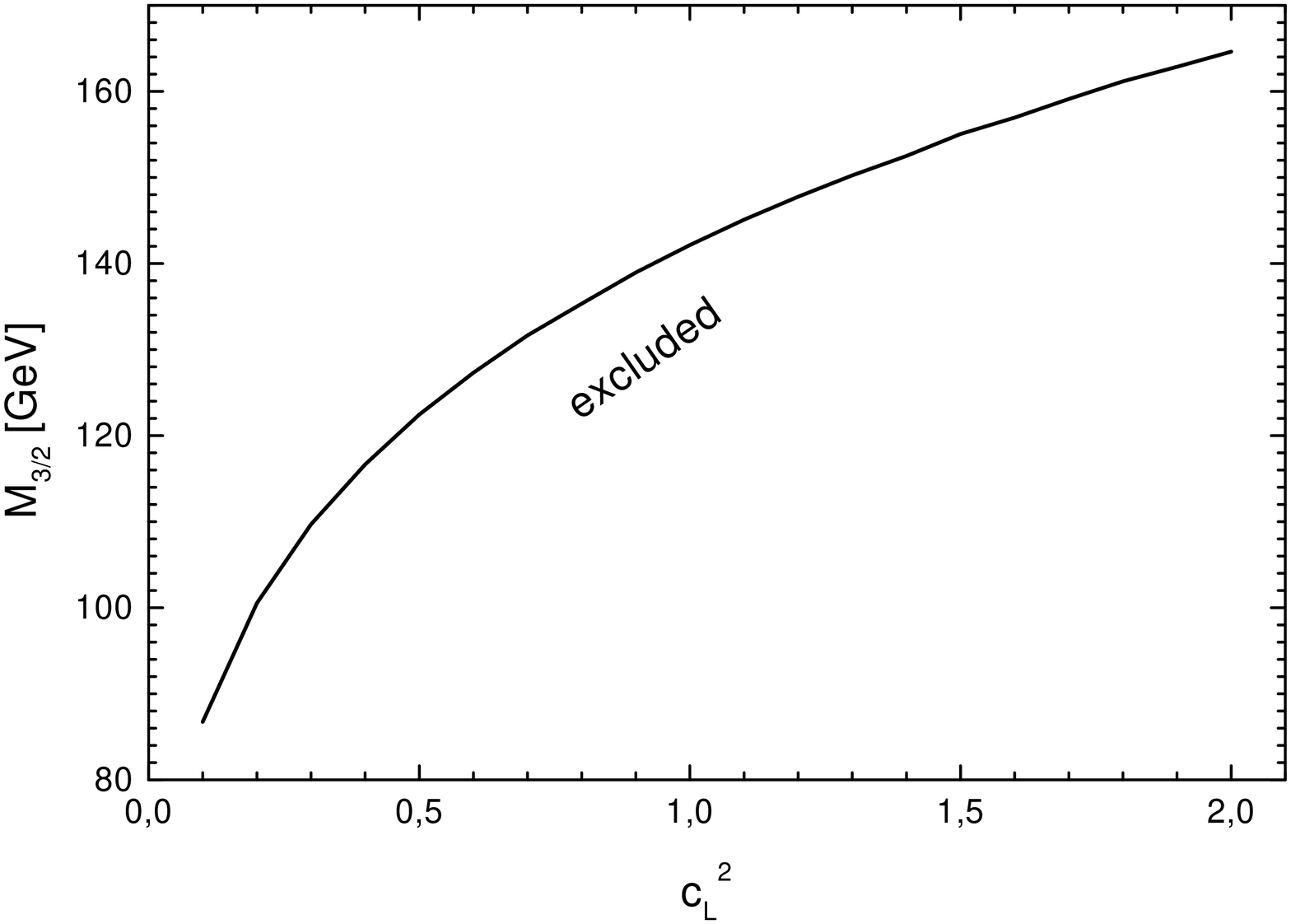,width=7.5cm,height=6cm}}
\end{center}
\caption{Same as Fig. 2 but for interaction ${\cal L}_{int}^{(2)}$.}
\label{fig3}
\end{minipage}
\end{figure}

We derived lower bounds on the exotic-lepton mass and couplings by a
$\chi^2$ fit, defining
$$
\chi^2(i) = \sum_{k} \Bigl( {\sigma^{(i)}_k - \sigma^{exp}_k \over \Delta 
\sigma_k} \Bigl)^2 \mbox{ , } i=1,2 \mbox{ ,}
$$
where $\sigma_k^{(i)} \equiv (d\sigma^{(i)}/d\Omega )_k$ denotes the 
theoretical value of the angular distribution for the $k^{th}$ bin, 
$\sigma_k^{exp} \equiv (d\sigma^{exp}/d\Omega)_k$ the corresponding 
experimental value measured by the OPAL Collaboration and $\Delta \sigma_k$ 
its associated experimental error for the $k^{th}$ bin. Bounds on
$M_{3/2}$ were computed for fixed values of the couplings. These lower
bounds at the $95\%$ confidence level correspond to an increase $\Delta
\chi^2 = 3.84$ with respect to the minimum. For $c_L^2 = 1$ and $c_R^2
= 0$, for instance, the lower limits are $M_{3/2} > 125\ GeV$ and $M_{3/2}
> 142\ GeV$ for interactions ${\cal L}_{int}^{(1)}$ and ${\cal L}_{int}^{(2)}$
respectively. Figs. \ref{fig2} and \ref{fig3} show the $95\%$ C.L. bounds on 
$M_{3/2}$ as
functions of $c_L^2$, with $c_R^2 = 0$. The lower limits are the same
if one interchanges $c_L$ and $c_R$.

\begin{figure}
\begin{minipage}{.470\linewidth}
\begin{center}
\mbox{\epsfig{file=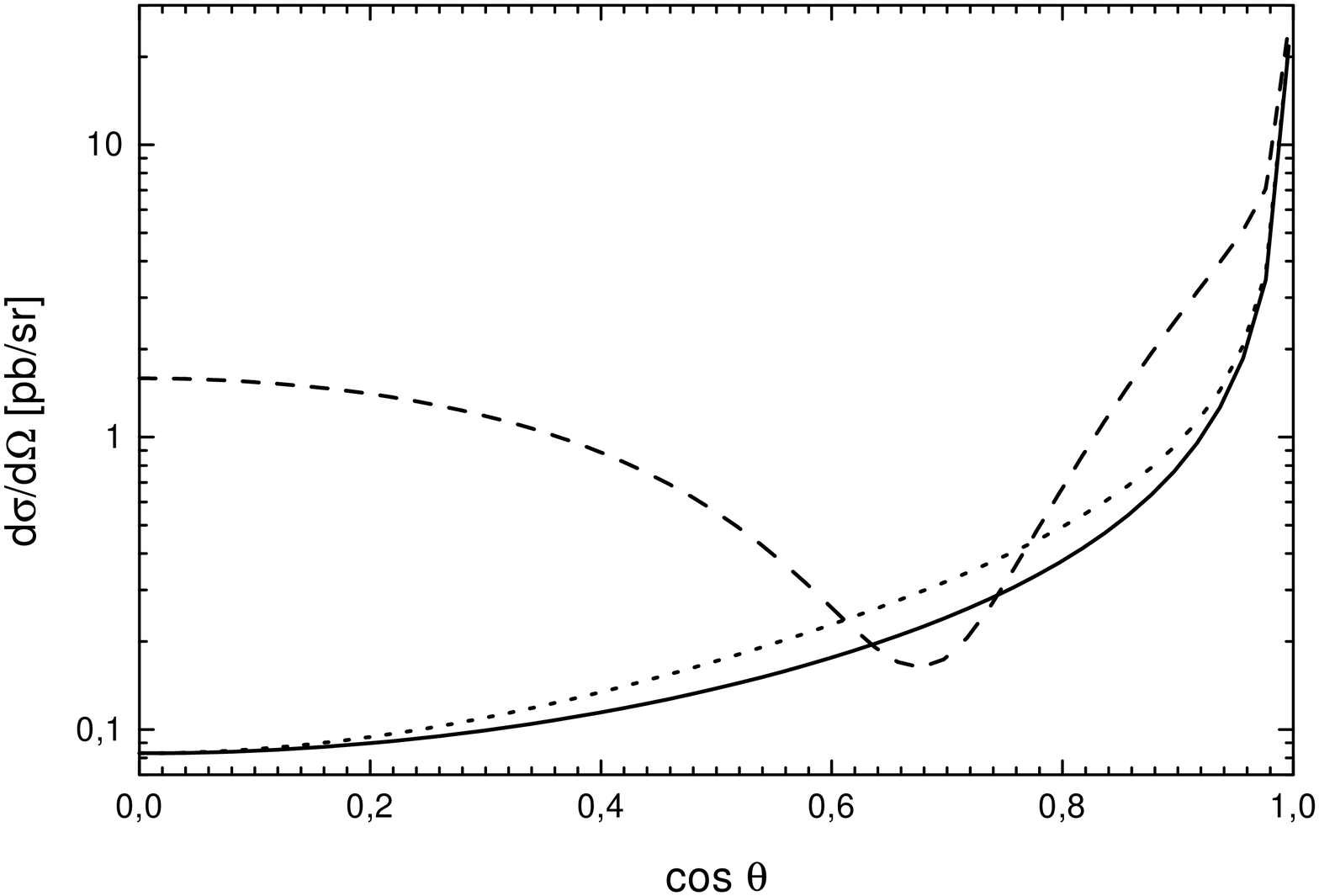,width=7.5cm,height=6cm}}
\end{center}
\caption{Angular distribution at $\sqrt s=500\ GeV$. The solid line 
represents the QED prediction, whereas the dashed (dottted) curve shows the 
total angular spectrum in the presence of the nonstandard interaction 
${\cal L}_{int}^{(1)}$, for an input mass $M_{3/2}=125\ GeV$ ($250\ GeV$).}
\label{fig4}
\end{minipage}
\nolinebreak
\hskip 0.5cm
\begin{minipage}{.470\linewidth}
\begin{center}
\mbox{\epsfig{file=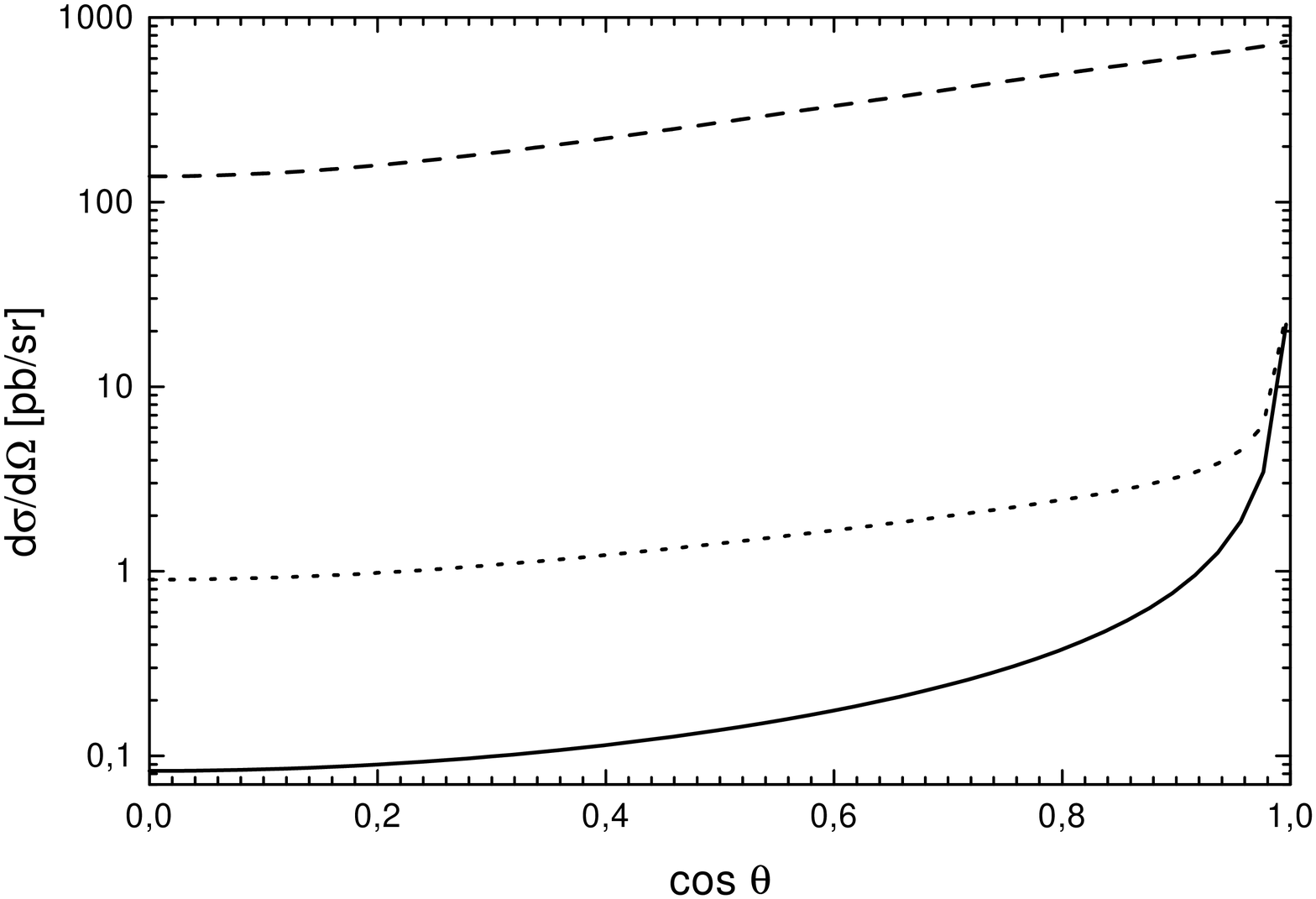,width=7.5cm,height=6cm}}
\end{center}
\caption{Angular distribution at $\sqrt s=500\ GeV$. The solid line 
represents the QED prediction, whereas the dashed (dottted) curve shows 
the total angular spectrum in the presence of the nonstandard interaction 
${\cal L}_{int}^{(2)}$, for an input mass $M_{3/2}=142\ GeV$ ($250\ GeV$).}
\label{fig5}
\end{minipage}
\end{figure}

\begin{figure}
\begin{minipage}{.470\linewidth}
\begin{center}
\mbox{\epsfig{file=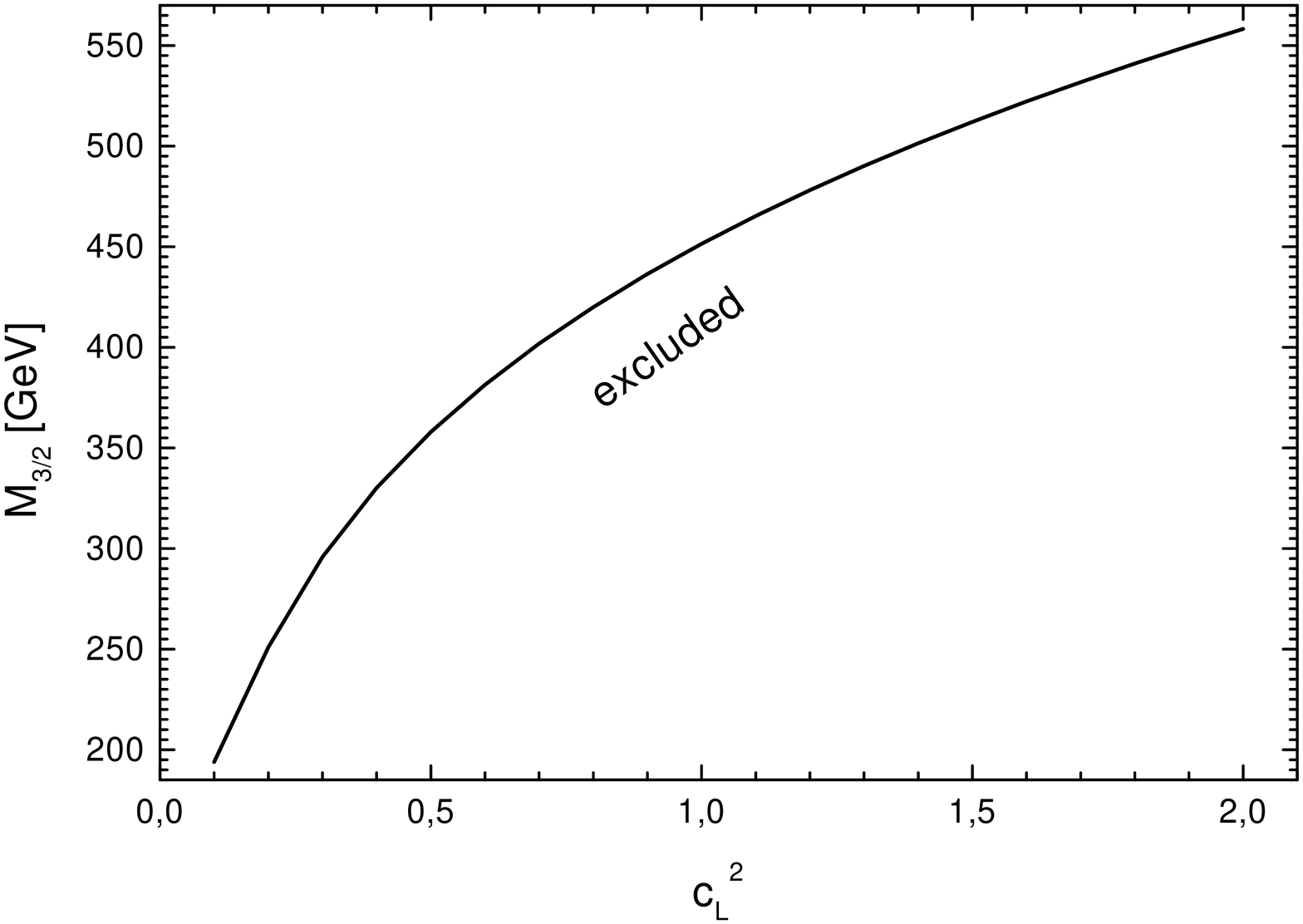,width=7.5cm,height=6cm}}
\end{center}
\caption{Same as Fig. 2 but for a NLC energy $\sqrt s=500\ GeV$.}
\label{fig6}
\end{minipage}
\nolinebreak
\hskip 0.5cm
\begin{minipage}{.470\linewidth}
\begin{center}
\mbox{\epsfig{file=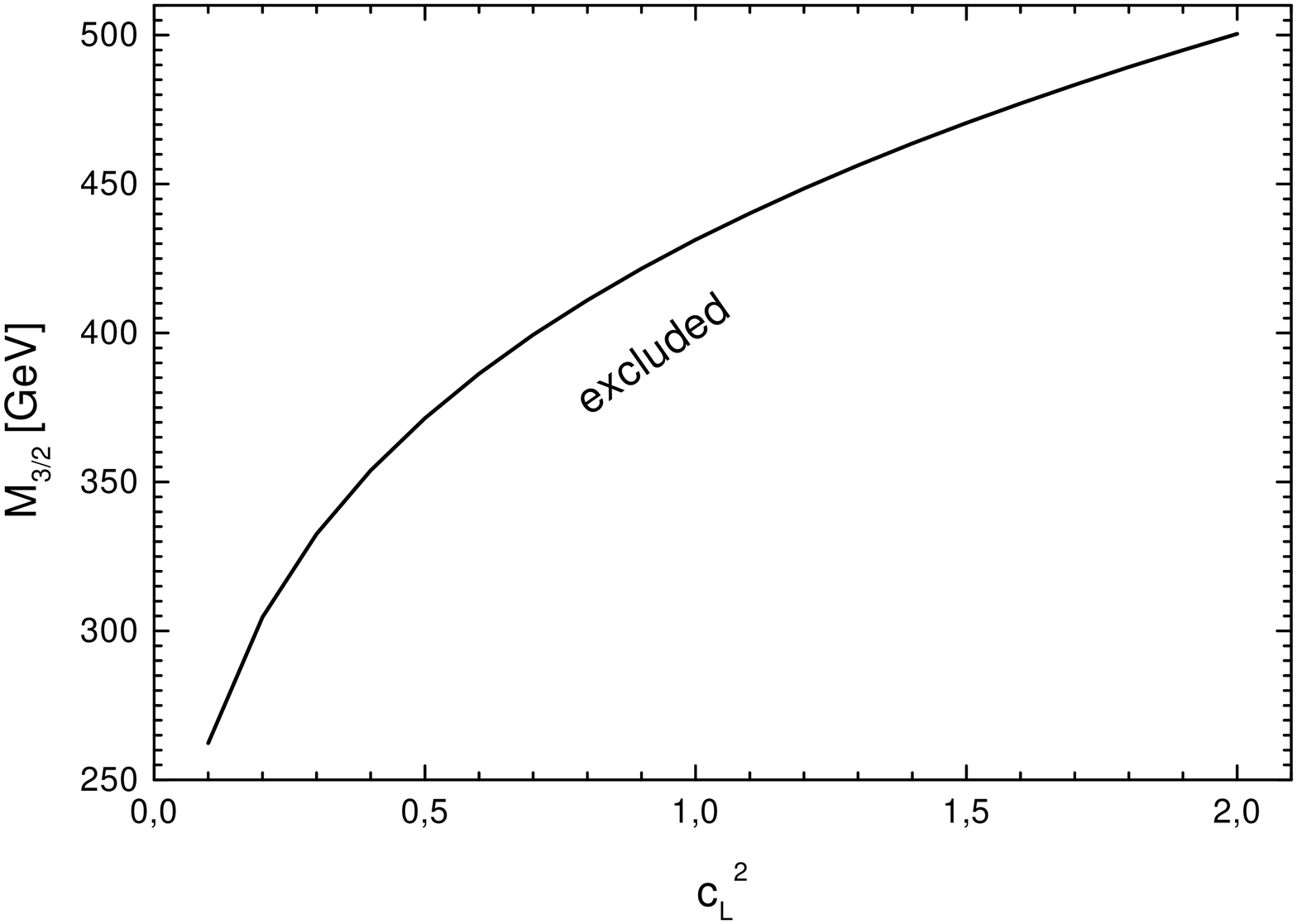,width=7.5cm,height=6cm}}
\end{center}
\caption{Same as Fig. 3 but for a NLC energy $\sqrt s=500\ GeV$.}
\label{fig7}
\end{minipage}
\end{figure}

The next generation of linear $e^+e^-$ colliders (NLC) will give
important contributions to the search of nonstandard physics. Angular
distributions for a $500\ GeV$ NLC are shown in Figs. \ref{fig4} and 
\ref{fig5}, considering
interactions ${\cal L}_{int}^{(1)}$ and ${\cal L}_{int}^{(2)}$ respectively, 
and assuming
an input mass $M_{3/2} = 250\ GeV$ or the lower bound which we obtained
from the OPAL data. We considred a cut in the polar angle $\theta$ such that 
$5^o < \theta < 175^o$. As expected, cross sections grow faster with energy in
the presence of the nonstandard interactions under discussion, the more so
in the case of ${\cal L}_{int}^{(2)}$, which contains a higher-dimensional 
operator. In order to estimate lower bounds in this case, we defined $\chi^2$ 
functions 
$$
\chi^2(i) = \sum_{k} \Bigl( {N(i)_k - N^{SM}_k \over \Delta 
N^{SM}_k} \Bigl)^2 \mbox { , } i=1,2 \mbox{ , }
$$
where $N(i)_k$ stands for the number of events in the
$k^{th}$ bin in the presence of the nonstandard electromagnetic
interactions, $N^{SM}_k$ the number of events predicted by the standard
model for the same bin, and $\Delta N^{SM}_k = \sqrt {N^{SM}_k +
(N^{SM}_k \delta)^2}$ the corresponding error, in which the
Poisson-distributed statistical error is combined in quadrature with the 
systematic error. We considered a conservative integrated luminosity of 
$10\ fb^{-1}$ and a typical systematic error $\delta = 2\%$ for a
measurement in a $500\ GeV$ NLC. The results of this $\chi^2$ analysis are
displayed in Figs. \ref{fig6} and \ref{fig7}. Clearly, the lower bounds can 
be considerably
improved by the experiments in the future $e^+e^-$ colliders.

We thank K. Sachs from OPAL for the data used in this paper. This work was 
partly supported by CNPq and FINEP.

\hrulefill

\end{document}